\documentclass[12pt,draftclsnofoot,onecolumn]{IEEEtran}
\usepackage {graphicx}
\begin{document}
%
\title{Soft Control on Collective Behavior of a Group of Autonomous Agents by a Shill Agent }
%
%

\author{Jing~Han,\,\,
        Ming~Li~\,
        and~Lei~Guo
\thanks{
        This work was supported by the National Natural Science Foundation of China.
        }
\thanks{Jing Han and Lei Guo are with the Institute of Systems Science, AMSS, Chinese Academy of Sciences,
        Beijing, 100080, China. Ming Li is with the Institute of Theoretical
        Physics, Chinese Academy of Sciences. Corresponding author: \texttt{hanjing@amss.ac.cn}.
        }
        }
%
%
%
\markboth{Published in Journal of Systems Science and
Complexity, 2006(19):54-62 }{Shell \MakeLowercase{\textit{et al.}}: Bare Demo of
IEEEtran.cls for Journals}
%



\maketitle

\begin{abstract}

This paper asks a new question: how can we control the collective
behavior of self-organized multi-agent systems? We try to answer
the question by proposing a new notion called `\textit{Soft
Control}', which keeps the local rule of the existing agents in
the system. We show the feasibility of \textit{soft control }by a
case study. Consider the simple but typical distributed
multi-agent model proposed by Vicsek \textit{et al.} for flocking
of birds: each agent moves with the same speed but with different
headings which are updated using a local rule based on the average
of its own heading and the headings of its neighbors. Most studies
of this model are about the self-organized collective behavior,
such as synchronization of headings. We want to intervene in the
collective behavior (headings) of the group by \textit{soft
control}. A specified method is to add a special agent, called a
`\textit{Shill}', which can be controlled by us but is treated as
an ordinary agent by other agents. We construct a control law for
the \textit{shill }so that it can synchronize the whole group to
an objective heading. This control law is proved to be effective
analytically and numerically. Note that \textit{soft control} is
different from the approach of \textit{distributed control}. It is
a natural way to intervene in the distributed systems. It may
bring out many interesting issues and challenges on the control of
complex systems.

\end{abstract}

\begin{keywords}
Collective Behavior, Multi-agent System, Soft Control, Boid Model,
Shill Agent
\end{keywords}

%
\IEEEpeerreviewmaketitle

\section{Introduction}
%
%
%
%

Collective behavior is the high level (macroscopic) property of a
self-organized system which consists of a large number of
(microscopic) individuals (agents). Examples are synchronization,
aggregation, phase transition, pattern formation, swarm
intelligence, fashion, etc. People found this kind of phenomena in
many systems, such as flocking of birds, schools of fish,
cooperation in ant colonies, panic of crowds \cite {panic}, norms
in economic systems \cite {special_agent_class_norm}, etc. Without
any question, collective behavior is one of the fundamental and
difficult topics of the study of complex systems. We classify the
research on collective behavior into three categories.

~~(I)  ~~{\bf {\it \textbf{Given the local rules of agents, what
is the collective behavior of the overall system?}}} Many people
have been working on this category which is about the mechanism of
how collective behavior emerges from multi-agent systems. ``More
is different''\cite{more_is_different}. The physicists have
applied theory of statistical physics to explored some simple
models, from the Ideal Gas model, Spin Glasses, to the panic model
and network dynamics.

~~(II) ~~{\bf {\it \textbf{Given the desired collective behavior,
what are the local rules for agents?}}} Some people work on this
category. One typical example is Swarm Intelligence. Since the
high level function of the overall system can be more than the sum
of all individuals, how do we construct robust intelligence by a
large number of locally interacting simple agents? An ant is
simple and often moves randomly. But a colony of ants can
efficiently find the shortest path from their nest to a food
source. This natural phenomenon inspired the Ant Colony
Optimization Algorithm\cite {swarm_intelligence}.

~~(III) ~~{\bf {\it \textbf{Given the local rules of agents, how
we control the collective behavior?}}} In some real applications,
it is very difficult or even impossible to change the local rules
of agents, such as the behavioral rules of people in panic and the
flying strategies of birds. Yet we need to control the system to
avoid danger or improve efficiency. Then what is the feasible way
to intervene in the collective behavior? There should be a way
especially for the multi-agent system that utilizes the property
of collective behavior. This question is not well noticed in the
control literature. In this paper, we propose a new control
notion, called `{\bf {\it \textbf{soft control}}}'\footnote{The
idea of {\it soft control} in this paper was inspired by
discussions at the {\it ``Systems, Control and the Complexity
Science'' Xiangshan Science Conference} (May, 2004, Beijing) and
the preliminary result was presented at {\it the 2nd
Chinese-Swedish Conference on Control} \cite
{our_Chinese-Swedish_conference}.}, which {\bf {\it \textbf{keeps
the basic local rule of the existing agents in the system}}}.
There are no global parameters can be adjusted to achieve the
control purpose, such as adjusting the temperature to change water
from liquid phase to gas/solid phase or broadcasting orders
directly to all agents by a controller. This feature makes a
difference between \emph{soft control} and the traditional
control.

These three categories are tightly related. The first category (I)
is about how a distributed system behaves and what emerges from
the system. Both of (II) and (III), however, require the system to
behave at the high level as what we expected, i.e., we know what
collective behavior should be emerged from the system. But (II)
focuses on how to design a distributed system by
constructing(changing) the local rules for interaction among
agents, and (III) focuses on how to intervene in the existing
system by a nondestructive way.

Since collective behavior is a kind of macroscopic feature of a
multi-agent system, usually adding one or a few more agents will
not affect it. But if those agents are special ones which can be
controlled, even though only a small number (in some cases, only
one), it can dramatically change the collective behavior as will
be shown in Section 3. So in this paper, a way to control the
system without changing the local rules of the existing agents is
to add one (or a few) special agent which can be controlled.
Therefore its behavior does not necessarily obey the local rules
as the ordinary agents do. However, the existing agents still
treat the special agent as an ordinary agent, so the special agent
only affects the local area with limited power. The special agent
is the only controlled part of the system and it indeed changes
the collective behavior of the system by `cheating' and 'seducing'
its neighboring ordinary agents. So we do not call the special
agent a leader. Instead, we call it a `{\bf {\it
\textbf{Shill}}}'.

{\it Soft control} is different from {\it distributed control}.
{\it Distributed control} recently gets more and more attention
because many real-world applications are distributed systems, such
as power networks and traffic systems. The system consists of many
interacting subsystems that all have their own controllers with
local feedback. The study of the relationship about performance
between the overall system and the subsystems falls into category
(I). On the other hand, designing a practical distributed system
for resource sharing and cooperation is a demanding and
complicated task. This research is of the category (II). So we can
see that both {\it soft control} and {\it distributed control}
concern with the macroscopic collective behavior (such as
synchronization) of the self-organized multi-agent system with
local rules. But in {\it distributed control,} every agent is
treated as a control system and has its control law (which is the
set of local rules); while in {\it soft control} framework, all
these agents are treated as one system and the control law is for
the {\it shill}. So {\it soft control} can be regarded as a way of
intervention in the distributed systems. With the growing
literature on complex systems, the challenge of controlling
complex systems is likely to become a key problem for control
scientists.

This paper is going to explore the idea of {\it soft control} by a
case study, where we will show how a {\it shill} can be used to
control the headings of a group of mobile agents demonstrated
based on the modified Boid model \cite {BOID} proposed by Vicsek
{\it et al} \cite {vicsek_model}. The current control theory
cannot be applied directly to this problem because our {\it soft
control} is a nondestructive intervention in a locally interacting
autonomous multi-agent system, which is not allowed to change the
local rules of the existing agents. The {\it formation control
}\cite {formation_control} mainly focuses on how to design the
local rules of a team of robots to maintain a geometric
configuration during movement, which is a very special kind of
collective behavior, and its approach belongs to {\it distributed
control}. The {\it pinning control} \cite {pinning_control} is
especially for stabilizing dynamical networks (with special
topology, such as random networks, small-world and scale-free
networks) by imposing controllers on a small fraction of selected
nodes in the network. No one appear to have directly studied the
above mentioned {\it soft control} problem in the literature.
Recently, Jadbabie et al \cite {jadbanaie} investigated a
multi-agent model which is slightly different from the one
proposed by Vicsek {\it et al}, they showed that under some { \it
a priori} connectivity conditions, the group will eventually move
in a same direction even without centralized control. This is a
problem of the category (I) of collective behavior study. An idea
in \cite{jadbanaie} worthy of mention is the {\it Leader
Following} model, which has a leader agent with fixed heading. But
it did not tell how to implement synchronization by a leader. The
so-called 'virtual leader'\cite{virtual_leader} is not treated as
an agent but part of the potential, which requires to introduce
new rules for the ordinary agents to recognize it.

The rest of this paper is organized as follows: The modified BOID
model proposed by Vicsek {\it et al} and the notion of {\it soft
control} are set up in Section 2. Section 3 is a case study for
{\it soft control}. We will solve a simple but non-trivial problem
by constructing an effective control law for the {\it shill}. Both
theoretical analysis and computer simulation show that one {\it
shill} is possible to turn the direction of the whole group. Some
concluding remarks are given in Section 4.

\section{Soft Control of the Modified Boid Model}

Eighteen years ago, Renolds proposed a simultaneous discrete-time
multi-agent model, which is called Boid\cite {BOID}, to catch the
natural phenomenon of flocking of birds and fish and make computer
animation. In this model, each bird decides its flying direction
only by looking at its current neighboring birds and using three
rules (\emph{Alignment}, \emph{Separation} and \emph{Cohesion})
based on the status of its neighbors. These rules are local and
simple but the overall system exhibits the flocking behavior. The
Boid model became very popular in complex systems studies.
Unfortunately, this simple model is not simple at all for
theoretical analysis.

In 1995, to investigate the emergence of self-ordered motion,
Vicsek {\it et al} \cite {vicsek_model} proposed a model which is
actually a modified version of Boid, by only keeping the {\it
Alignment} rule{\bf --} steer towards the average heading of
neighbors, which still can catch the flocking behavior. There are
$n$ agents $(x_i(\cdot),\theta_i(\cdot))$ for $n$ birds, labelled
from $1$ through $n$, all moving simultaneously in the
two-dimensional space. The velocity of agent $i$ at time $t$ is
defined by
\begin{equation} \label{velocity}
 \emph{\textbf{v}}_{i}(t)=(v\cos(\theta_{i}(t)),
v\sin(\theta_{i}(t)))
\end{equation}
which is constructed to have an absolute value $v$ and a heading
given by the angle $\theta_{i}(t)\in[0,2\pi)$. The position of
agent $i$ at time $t$ is denoted as $\textbf{\emph{x}}_{i}(t)$.
The neighborhood of agent $i$ at time $t$ is defined as
$N_{i}(t)=\{j|\|\emph{\textbf{x}}_{j}(t)-\emph{\textbf{x}}_{i}(t)\|\leq
r, j=1,2,\ldots,n\}$, where $r$ is the radius of the neighborhood
circle. For any agent $i$, its heading and position are updated by
(\ref{heading_update}) and (\ref{position_update}) below if
ignoring noise:
\begin{equation} \label{heading_update} \theta
_{i}(t+1)= \langle \theta _{i}(t)\rangle_{r}
\end{equation}
\begin{equation}
\label{position_update} \emph{\textbf{x}}_{i}(t+1)
=\emph{\textbf{x}}_{i}(t) + \emph{\textbf{v}}_{i}(t+1)
\end {equation}
where $\emph{\textbf{v}}_{i}(t+1)$ is defined by (\ref{velocity}),
and $\langle\theta _{i}(t)\rangle_{r }$ is the angle of the sum of
the velocity vectors of neighbors of agent $i$:$\sum\limits_{j\in
N_i(t)}{\emph{\textbf{v}}_{j}(t)}$. In other words, $\langle
\theta _{i}(t)\rangle_{r}$ can be obtained by

\begin{equation}\label{avg_heading}
\arctan(\sum\limits_{j \in N_i (t)} {\sin(\theta _j (t))}/
\sum\limits_{j \in N_i (t)} {\cos (\theta _j (t))})
\end{equation}with some necessary regulations reflecting angles of $[0, 2\pi )$.

Some recent results of Jadbabaie et al \cite {jadbanaie} have
shown that under some connectivity conditions, the group will
eventually move in a same direction. However, the model of \cite
{jadbanaie} calculates $\langle \theta _{i}(t)\rangle_{r}$ by
simply taking the average of angles of all neighbors. This brings
convenience for mathematical analysis, but the system will
sometimes exhibit counter-intuitive phenomena as they claimed in
their paper\footnote{ This is because the zero angle is special
than other angles in $(0, 2\pi)$. The angles do not symmetrically
behave. }. So this paper keeps the updating rule of the model
proposed by Vicsek {\it et al}, even though the analysis result
about synchronization of the model in \cite {jadbanaie} cannot be
utilized\footnote{ The result about synchronization for the model
of \cite {jadbanaie} is not true for the model of Vicsek {\it et
al}. Here is a counterexample: when $v=0$, six agents are
regularly put on the edge of a circle with radius of $r$, which
form an ordinary hexagon. So each agent has three neighbors
(because itself is also counted). The headings are $0, \pi/3,
2\pi/3, \pi, 4\pi/3, 5\pi/3$. This case will synchronize for the
model in \cite {jadbanaie}. But it will not for the model we
consider here, i.e. the Vicsek \emph{et al} model without noise.}.

One can think of many related questions: what is the way to help
the group to synchronize if the group will not synchronize by
self-organization? What if the group self-organize to a
synchronized direction $\alpha $ which is not what we want? What
if we want the group to fly to a desired destination? And so on.

In this paper, we consider the case of adding one {\it shill} to
control the collective behavior with the {\it shill} denoted as $(
\emph{\textbf{x}}_{0}, \theta _{0})$. The ordinary agent
($i=1,\ldots,n$) still keep the local rule as formula
(\ref{heading_update})-(\ref{position_update}) to update its
heading and position. The only difference is that the neighborhood
$N_{i}(t)$ for agent $i$ at time $t$ will consider the {\it
shill}:
 $N_{i}(t)=\{j|\|\emph{\textbf{x}}_{j}(t)-\emph{\textbf{x}}_{i}(t)\|\leq
r, j=\textbf{0},1,2,\ldots,n\}$. And the {\it shill} does not
exactly obey the local rules as the ordinary agents do, its
position and heading is decided by the control law {\bf {\it u}}:

\begin{center}
$( \emph{\textbf{x}}_{0}(t), \theta _{0}(t) )= u
(\emph{\textbf{x}}_{1}(t), \ldots , \emph{\textbf{x}}_{n}(t),
\theta _{1}(t), \ldots , \theta_{n}(t), t)$
\end{center}
Note that $( \emph{\textbf{x}}_{0}, \theta _{0})$ may be subject
to some constraints. For example, the constraint of {\bf {\it
x}}$_{0}$(t+1) = {\bf {\it x}}$_{0 }$(t) + {\bf {\it v}}$_{0
}(t+1)$ will make the {\it shill} behave like an ordinary agent
except the way it decides its own heading. Without any constraint,
the {\it shill} can fly to anywhere with a `deceptive' heading.
\vskip 0.5cm

 \centerline{
 \scalebox{0.50}{\includegraphics{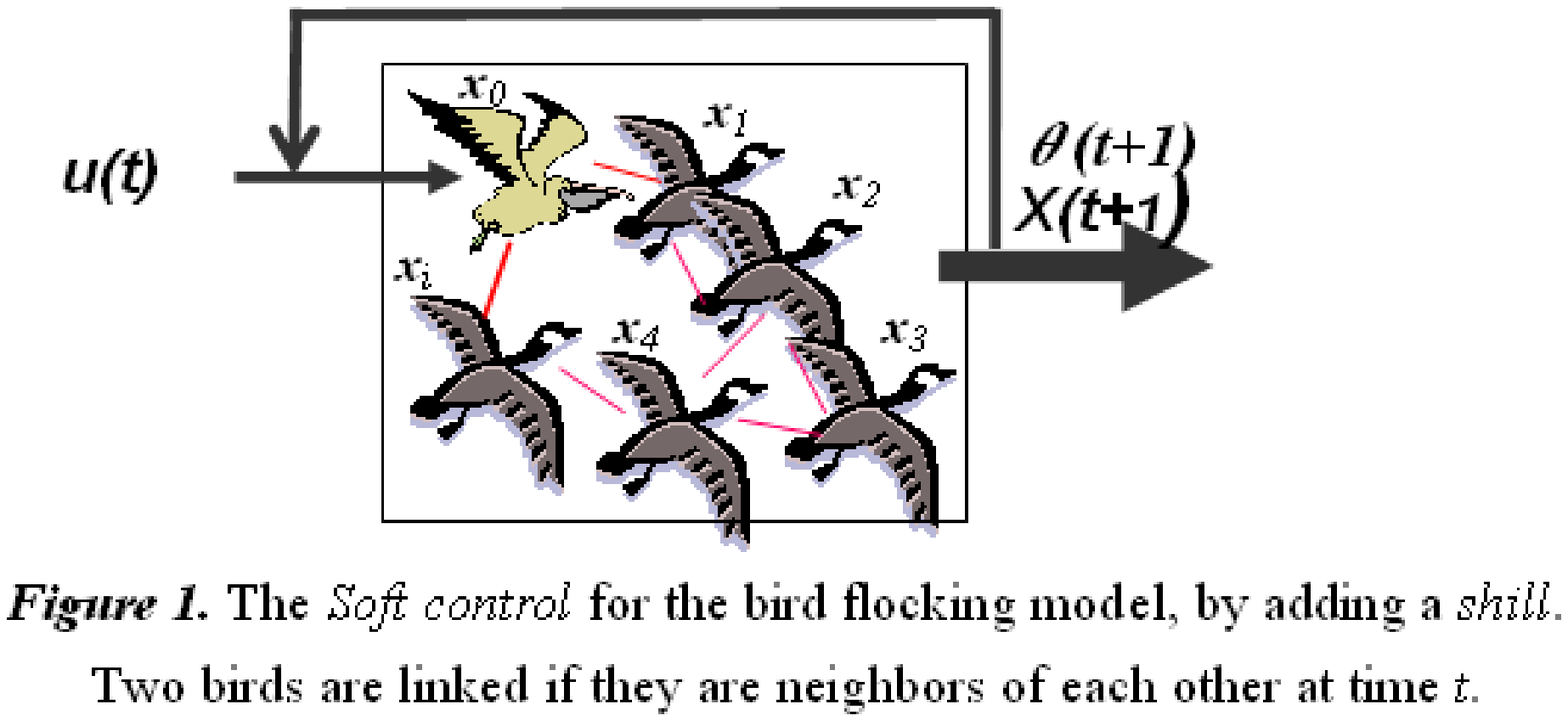}}
\label{fig1} }


The control law {\bf {\it u}} will be different for different
control purposes/tasks, such as to synchronize headings of the
group ($n$ agents), to keep the group connected or to dissolve a
group, to turn headings of the group (minimal circling), to lead
the group to a destination (in a shortest time), to avoid hitting
an object, etc.

There are many problems to be considered in terms of control:
under what conditions we can use the {\it shill} to control the
state of the multi-agent system (the controllability problem);
which kind of state information can be used for control (the
observation problem), e.g., the {\it shill} can only see some
nearest agents but not all other agents; how can one design the
control law when the local rule and other information is not clear
(the learning and adaptation problem), etc. We do not intend to
answer all these questions in this paper. As a beginning, we will
just show how it works for a specific case, which is to
synchronize and turn the headings of a group.

\section{A Case Study }
Flocking of birds in airports is dangerous. How we drive them
away? In this case, obviously we can not use centralized control,
such as broadcasting orders to the birds to change to the desired
flying direction. The current solution is to use cannon to shoot
them away. But can we drive them away by {\it soft control}? If
the modified Boid model proposed by Vicsek {\it et al} works for
real birds flocking, and if we can make a controllable powerful
robot bird, can we use this robot bird as the {\it shill} and
guide flight of the birds? In this section, we will study a simple
but non-trivial related problem \footnote{In flocking behavior of
real animals, more factors such as different sensory systems
should be considered. The Boid model is an abstract model which
successfully simulates the flocking behavior. But no one has
proved it to be true for real birds. So in this paper we just want
to use a simple model to demonstrate how \emph{soft control} works
for self-organized multi-agent systems in general, but not to
solve the specified airport birds problem.}.

The {\it problem }we consider here is: for a group of $n$ agents
with initial heading of $\theta _{i}(0)\in [0, \pi )$, $1\leq
i\leq n$, what is the control law for the {\it shill}, so that all
agents will move to the direction of $\pi$ eventually?

Suppose the local rule about the ordinary agents is known. Suppose
also that the position $x_0(t)$ and heading $\theta_0(t)$ of the
{\it shill} can be controlled at any time step $t$. Suppose
further that the state information (headings and positions) of all
ordinary agents are observable at any time step. Now we propose an
effective control law $u_{\beta }$ which is defined as (see Fig.
2):

\vskip 0.5cm

\centerline{\scalebox {0.52}{\includegraphics{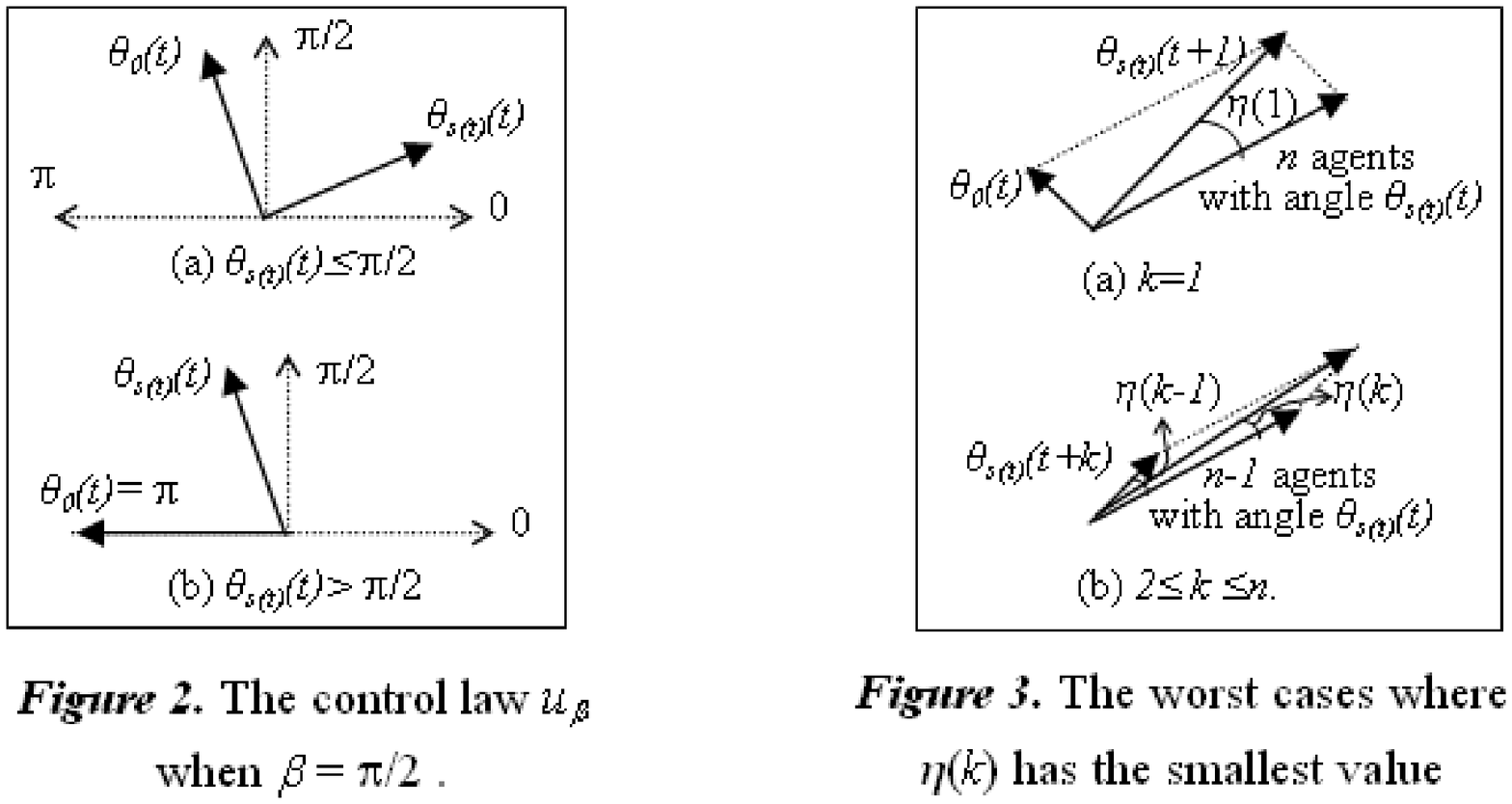}}}
%
\label{fig23}

\begin{equation} \label{controlaw_position} x_{0}(t) =
x_{s(t)}(t)
\end {equation}
\begin{equation} \label{controlaw_heading}
\theta_{0}(t)= \left\{
\begin{array}{cl}
\theta_{s(t)}(t)+ \beta \quad  & \mbox {if} \quad \theta_{s(t)}\leq \pi-\beta;\\
\pi\quad  & \mbox {if} \quad \theta_{s(t)} > \pi-\beta;\\
\end{array}
\right.
\end {equation}where $\beta \in (0, \pi )$ is a constant, and $s(t)$ is
defined as the 'worst' agent (or one of the 'worst' agents):
$s(t)=\mathop {argmin}\limits_{1 \le i \le n} \{\theta _i (t)\}$;

\vskip 0.5cm

The intuition behinds $u_{\beta }$ is to put the {\it shill} to
the position of the `worst' agent $s(t)$ at each time step $t$ and
try to `pull' it to the desired direction $\pi $. This greatly
reduces the search space of positions and simplifies the control
strategy. Note that we can not use $\theta _{0}(t)=\pi$ for all
the time because: (\textbf{A}) in the case where $\theta
_{s(t)}(t)=0$ and the all the neighboring ordinary agents of the
\emph{shill} have headings of \emph{zero} degree, the {\it shill}
with $\theta _{0}(t)=\pi $ can not change $\theta _{s(t)}(t)$
according to the update rule (\ref{heading_update}); (\textbf{B})
{to avoid the ill-case where both of the numerator and denominator
in (\ref{avg_heading}) are \emph{zero}\footnote{In the ill-case,
$\sum\limits_{j \in N_i (t)} {\sin(\theta _j (t))}=0$ implies
$\theta_j(t)=0$ or $\pi$ for all $j\in N_i (t)$ because $\sin
\alpha >0$ for $\alpha\in(o,\pi)$. Consequently, by
$\sum\limits_{j \in N_i (t)} {\cos(\theta _j (t))}=0$ we further
conclude that half of the headings should be \emph{zero} and the
other half should be $\pi$. So there must be at least one agent
$j\neq0$ such that $\theta_j(t)=0$ (since $\theta_0(t)>0$ for
$t>0$ by the control law), which means $\theta_{s(t)}(t)=0$, and
so we must have $\theta_0(t)=\beta\in{(0,\pi)}$ if $0\in N_i(t)$.
However, this will contradict with the fact that $\theta_j(t)=0$
or $\pi$ for all $j\in N_i (t)$. On the other hand, if $0\notin
N_i(t)$, $\sum\limits_{j \in N_i (t)} {\sin(\theta _j (t))}=0$
only holds when $\theta_j(t)=0$ for all $j\in N_i (t)$ since
obviously $\theta_j(t)<\pi$ for all $j=1,\ldots,n$ and $t\geq0$,
but in this case we have $\sum\limits_{j \in N_i (t)} {\cos(\theta
_j (t))}>0$. So the ill-case will never happen in
(\ref{avg_heading}). }. But in the final stages we do need $\theta
_{0}=\pi $ to help the system to synchronize.


In the sequel, we use $\Delta(t) = \pi -\theta _{s(t)}(t)$ to
denote the distance between the angel of the `worst' agent and the
objective angel $\pi $ at time $t$. We now have the following main
technical result of the paper:

{\bf {\it \textbf{Theorem 1}}}. For any $n \ge 2$, $r\ge 0$, $v\ge
0$, and any initial headings and positions $\theta _{i}(0)\in [0,
\pi)$, $x_i(0)\in R^{2}$, $(1\le i \le n)$, the update rule
(\ref{heading_update})-(\ref{position_update}) and the {\it soft
control} law $u_{\beta }$ defined by
(\ref{controlaw_position})-(\ref{controlaw_heading}) will lead to
the asymptotic synchronization of the group, i.e.,
$\lim_{t\rightarrow \infty} \Delta(t)=0$.

{\bf {\it \textbf{Proof.} }} Because $\pi \ge \theta _{0}(t)>
\theta_{s(t)}(t)\geq0$, and $\pi
> \theta_{i}(t)\ge \theta _{s(t)}(t)$ for $1\le i \le n$, it is
obvious by the update rule (\ref{heading_update}) that $\pi>\theta
_{i}(t+1)\ge \theta _{s(t)}(t)$ for $1 \le i \le n$ which implies
that $\theta _{s(t)}(t)\le \theta _{s(t + 1)}(t+1) $. Hence by
definition of $\Delta(t)$, we have $\Delta(t)\ge \Delta(t+1)\geq
0$, so $\Delta(t)$ will converge to some limit $\mu \ge 0$. If
$\mu>0$, then for any $0<\varepsilon< \mu$, we know that
$\Delta(t) \ge \varepsilon
>0$ holds for all large $t$.Consequently, by letting   $t\to\infty$ in the following Lemma 2, we have $\mu \le \mu-\delta$ which
contradicts with $\delta>$0.So we can conclude $\mu=0$. This
completes the proof of Theorem 1.

\vskip 0.5cm
 The main task now is to prove the following lemma.

{\bf {\it \textbf{Lemma 2}}}{\bf . }If for some $t>0$ and
$\varepsilon > 0$, $\Delta(t) \ge\varepsilon>0$, then there exists
a constant $\delta \in (0,\varepsilon)$ which does not depend on
$t$, such that $\Delta(t+n)\le\Delta(t)-\delta$.

{\bf {\it \textbf{Proof.}}} ~~ At any time step $t$, the {\it
shill} is affecting the agent $s(t)$ with heading $\theta_{0}(t)$.
We now proceed to analyze the lower bound to the angle change
$\theta _{s(t)}(t+k)-
 \theta_{s(t)}(t)$ for $1 \le  k \le   n$, which is
denoted by $\eta (k)$. For $k=1$, the {\it shill} is affecting the
agent $s(t)$, so the worst case where $\eta (1)$ has the smallest
value is that the agent $s(t)$ is surrounded by $n-1$ agents with
heading $\theta _{s(t)}(t)$ (see Fig.3-(a)). For $k \ge 2$,
however, the {\it shill} may not affect the agent $s(t)$, so the
worst case where $\eta (k)$ has the smallest value is that the
agent $s(t)$ is surrounded by $n-1$ agents with heading $\theta
_{s(t)}(t)$ again, but without the {\it shill} in its neighborhood
(see Fig.3-(b)). Now by the definition of $\eta (k)$ we have
\begin{equation}
\theta _{s(t)}(t+k) \ge   \theta _{s(t)}(t)+
 \eta (k),
\end{equation}
\noindent and a simple calculation will give
\begin{equation} \label{cc}
\eta(k)= \left\{
\begin{array}{cl}
\min \{\arctan (\sin \beta /(n+\cos\beta )), \arctan(\sin \varepsilon /(n+\cos \varepsilon )) \}, \quad k=1;\\
\arctan(\sin(\eta (k-1))/(n-1 +\cos(\eta (k-1))), \quad \quad \quad \quad 2 \le k  \leq n;\\
\end{array}
\right.
\end {equation}

\vskip 0.5cm

First, note that  $\Delta (t) \ge  \varepsilon >0$ and the
function $\psi (\alpha )= \arctan(\sin \alpha / (m +\cos \alpha
)), (m>0)$ is monotonically increasing when $\alpha  \in [0,
\arccos(-1/m)]$ and monotonically decreasing when $\alpha \in
[\arccos(-1/m), \pi]$.

Second, to explain the definition of $\eta(1)$ in (\ref{cc}) we
need to show that
\begin{equation}
\label{eta}
 0<\eta(1) \leq arctan(\sin(\theta_{0}(t)-\theta_{s(t)}(t))/(n+\cos(\theta_{0}(t)-
\theta _{s(t)}(t)))).
\end{equation}

In fact, by the control law (\ref{controlaw_heading}) and the
expression of $\eta(1)$ in (\ref{cc}), the above inequality
(\ref{eta}) is trivial in the case where $\theta_{s(t)}\leq
\pi-\beta$. So we only need to consider the case where
$\theta_{s(t)} > \pi-\beta$. In this case the control law is
$\theta_0(t) = \pi$. We consider two subcases as follows:

(i) $\theta_0(t) - \theta_{s(t)}(t) \geq \arccos(- {1\over n})$.
Since $\theta_0(t)- \theta_{s(t)}(t) \leq \pi -(\pi-\beta) =
\beta$, by the monotonicity of the function $\psi(\cdot)$, we know
that the claim (\ref{eta}) is true.

(ii) $\theta_0(t) - \theta_{s(t)}(t) < \arccos(- {1\over n})$.
Since $\theta_0(t) - \theta_{s(t)}(t)= \Delta(t) \geq
\varepsilon$, by the monotonicity of the function $\psi(\cdot)$
again, we know that the claim (\ref{eta}) is also true.

Hence, (\ref{eta}) holds in any case.

\vskip 0.5cm

Now, note that $0 < \eta (1) <\pi/2$, and again by the
monotonicity of the function $\psi(\cdot)$,  we have $0< \eta (n)
<\eta(n-1)<\ldots < \eta ( 1)<\pi/2$.

Let us take $\delta =\eta (n)$ which is a finite positive value
depending on $n$, $\beta$ and $\varepsilon $ only. Obviously,  we
have $\theta _{s(t)}(t+n)-\theta _{s(t)}(t)  \ge
 \delta $.
 \vskip 0.4cm

Next, we consider the other agents: $i =1,2,\ldots,n, i \ne s(t)$.

Let  $\Lambda (k)$ denote the set of agents whose angles are
inside $(\theta _{s(t)}(t), \theta _{s(t)}(t)+\delta )$ at time
step $t+k$, i.e.,

$\Lambda (k)=\{i|\theta _{s(t)}(t)< \theta _{i}(t+k)<\theta
_{s(t)}(t)+\delta,\,\, i =1,2,\,\ldots,n \}.$

We now proceed to show that we must have $\Lambda (n)=\emptyset$.

First of all, as shown above, $s(t)$ does not belong to $\Lambda
(k)$ for $1 \le k \le n$ because $\theta _{s(t)}(t+k) \ge  \theta
_{s(t)}(t)+  \eta (k) \ge \theta _{s(t)}(t)+ \delta $. So we have
$\vert \Lambda (k)\vert \le
 n-1$ for $1 \le k \le n$.

Therefore, at time step $t+1$, there are at most $n-1$ agents with
angles which are less than $\theta _{s(t)}(t)+\delta $. If
$\Lambda (1) \ne \emptyset $, $s(t+1)$ will be picked up from
$\Lambda (1)$. The {\it shill} will change the heading of agent
$s(t+1)$. Because $\theta _{s(t + 1)}(t+1) \ge   \theta
_{s(t)}(t)$ and $\theta _{s(t)}(t+k) \ge
 \theta _{s(t)}(t)+\eta (k)$ for any $t$, we have for $k \geq 2$,

$\theta _{s(t + 1)}(t+k) \ge \theta _{s(t + 1)}(t+1)+\eta (k-1)
\ge \theta_{s(t + 1)}(t+1)+  \delta  \ge   \theta _{s(t)}(t)+
\delta $.

So $s(t+1)$ will not belong to $\Lambda (k)$ for any $k \in
[2,n]$, we have $\vert \Lambda (k)\vert  \le  n-2$ for $2 \le k
\le n$. By repeating this argument, we get for $1\leq d \le k \le
n$, $\theta _{s(t + d)}(t+k) \ge \theta _{s(t)}(t) + \delta$  and
$\vert \Lambda (k)\vert \le n-d$. So by taking $d = k = n$, we
have $\vert \Lambda (n)\vert =0$. This means that after at most
$n$ steps, $\Lambda $ will be empty.

Consequently, we know that $\theta_i(t+n) \geq
\theta_{s(t)}(t)+\delta$ for all $i$. Then we get
$\theta_{s(t+n)}(t+n) \geq \theta_{s(t)}(t)+\delta\Rightarrow
\Delta (t+n) \le   \Delta (t)-\delta $. This complete the proof of
Lemma 2.
 \vskip 0.5cm

The snapshots of the relating computer simulation are showed in
Figure 4. The video of demo and the simulation program can be
downloaded from the project website \cite {our_project_home}.

Although our case study is a simple starting point, it shows that
it is possible to change the collective behavior (headings) of a
group by {\it soft control}. Based on this work, we are going to
explore more complicated control problems, such as what if the
energy of the {\it shill} is concerned (the energy of the shill is
limited so that it can not 'jump' as in
(\ref{controlaw_position})), what if the {\it shill} only can see
locally, and what if the {\it shill} also needs to keep the group
connected while turning the group, etc.

To support this research, we have put a computer demo for {\it
soft control} of the modified Boid model which is like a computer
game on the project homepage \cite {our_project_home}. After
initially putting $n$ agents on the two-dimensional space, the
user can manually control the position and heading of the {\it
shill} by keyboard and mouse. It is an open competition for the
best control strategy.
 \vskip 0.5cm
\centerline{ \scalebox {0.75}{\includegraphics {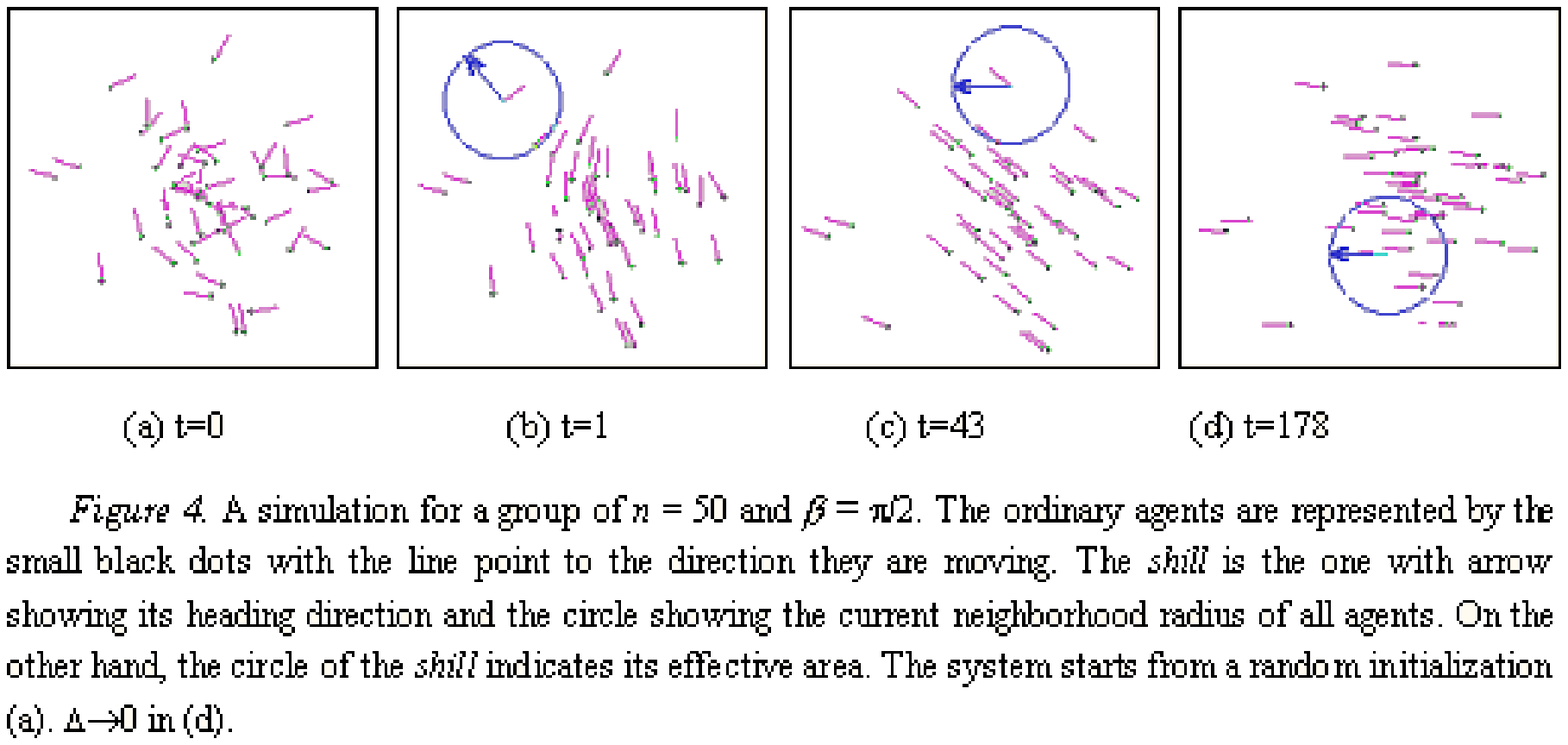}}}


\section{Concluding Remarks}

The collective behavior of complex systems is the macroscopic
feature that we concerned and studied in this paper. How we
control or intervene in a multi-agent system without changing the
local rule of the existing agents is an important issue but has
not been well recognized yet. The idea of {\it soft control } that
we introduced and explored here appear to be an initial attempt to
tackle this issue. We believe that our general \emph{soft control}
framework reflects a large class of control problems in complex
systems.

In this paper,  we have restricted ourselves on the study of the
modified Boid model which does catch certain key properties of
many real world complex systems (see,
e.g.,\cite{vicsek_model}\cite{BOID}\cite{jadbanaie}). Without
changing the local rule of the existing agents, we have added a
{\it shill} in the distributed system and designed the control law
for the {\it shill} carefully for the control purpose. Although
the {\it shill} does not behave as the ordinary agents do, it is
not recognized by the ordinary agents and is therefore still
treated as an ordinary agent by them. This {\it soft control }idea
works well as we have shown in the case study. We would like to
emphasis that the idea of adding a {\it shill} is just one of the
methods of {\it soft control}, and for different systems, they
would be certainly different.

There are many potential applications of {\it soft control}.
Currently a project called {\it Claytronics} \cite {claytronics},
a new form of programmable matter, is to study how to form
interesting dynamic shapes and configurations by simple and local
interactions of a billion micro-scale units. This research is in
the category (II) of collective behavior. Using {\it soft control}
to help to form shapes may release us from designing subtle local
rules. Another possible case is to use {\it soft control} in the
language diffusion model to guide evolution of language, and it
might be able to save dying language \cite {language}. A recent
study in economics \cite {special_agent_class_norm} shows that a
small proportion of special agents who have preferences can
dramatically alter the expected waiting time between norm
transitions in a population of agents who repeatedly play the Nash
demand game. If those special agents have feedbacks, how does the
{\it soft control} help changing the transition of norms more
efficiently ? How we use {\it soft control} to avoid and handle
panic of crowd \cite {panic} ... In conclusion, we need a theory
of {\it soft control} to efficiently intervene in many complex
systems.

\section*{Acknowledgment}
The first author would like to thank {\it Prof.} {\it John
Holland}, {\it  Dr. Liu Zhixin} and {\it Prof. Eric Smith }for
helpful discussions.

\end{document}